\documentclass[journal,comsoc]{IEEEtran}
\usepackage[T1]{fontenc}

\ifCLASSINFOpdf
\else
\fi
\usepackage{amsmath}
\usepackage[cmintegrals]{newtxmath}
\usepackage{array}
\usepackage{url}
\usepackage{subcaption}
\usepackage{graphicx}
\newcommand{\RNum}[1]{\uppercase\expandafter{\romannumeral #1\relax}}
\DeclareMathOperator{\vect}{vec}
\DeclareMathOperator*{\E}{\mathbb{E}}
\usepackage[T1]{fontenc}
\usepackage[utf8]{inputenc}
\usepackage{pgfplots}
\usepackage{grffile}
\pgfplotsset{compat=newest}
\usetikzlibrary{plotmarks}
\usetikzlibrary{arrows.meta}
\usepgfplotslibrary{patchplots}
\usepackage{amsmath}
\usepackage{pgfplots}
\usepackage{grffile}
\pgfplotsset{compat=newest}
\usetikzlibrary{plotmarks}
\usetikzlibrary{arrows.meta}
\usepgfplotslibrary{patchplots}
\usepackage{amsmath}
\usepackage{pgfplots}
\usepgfplotslibrary{groupplots}
\usetikzlibrary{fit}

\definecolor{mycolor1}{rgb}{0.00000,0.44700,0.74100}%
\definecolor{mycolor2}{rgb}{0.85000,0.32500,0.09800}%
\definecolor{mycolor3}{rgb}{0.92900,0.69400,0.12500}%
\definecolor{mycolor4}{rgb}{0.49400,0.18400,0.55600}%
\definecolor{mycolor5}{rgb}{0.46600,0.67400,0.18800}%
\definecolor{mycolor6}{rgb}{0.30100,0.74500,0.93300}%
\definecolor{mycolor7}{rgb}{0.63500,0.07800,0.18400}%
\definecolor{mycolor8}{rgb}{0,0.749019608,1}%
\definecolor{mycolor9}{rgb}{0,0,1}%
\definecolor{mycolor10}{rgb}{0.596078431,0.960784314,1}%
\definecolor{mycolor11}{rgb}{0,0.960784314,1}%
\definecolor{mycolor12}{rgb}{0.678431373,1,0.184313725}%
\definecolor{mycolor13}{rgb}{0,0.780392157,0.549019608}%
\definecolor{mycolor14}{rgb}{1,0.756862745,0.756862745}%
\definecolor{mycolor15}{rgb}{1.00000,0.00000,1.00000}%
\usepackage{pgfplots}

\newenvironment{customlegend}[1][]{%
    \begingroup
    \csname pgfplots@init@cleared@structures\endcsname
    \pgfplotsset{#1}%
}{%
    \csname pgfplots@createlegend\endcsname
    \endgroup
}%
\def\addlegendimage{\csname pgfplots@addlegendimage\endcsname}

\newcommand{\addlegendimageintext}[1]{%
    \tikz {
        \begin{customlegend}[
            legend entries={\empty},
            legend style={
                draw=none,
                inner sep=0pt,
                column sep=0pt,
                nodes={inner sep=0pt}}]
        \addlegendimage{#1}
        \end{customlegend}
    }%
}
\begin{document}
	%
	\title{Deep Learning for DOA Estimation in MIMO Radar Systems via Emulation of Large Antenna Arrays}
	%
	%
	%
	
	\author{Aya Mostafa Ahmed\IEEEauthorrefmark{1}
,~\IEEEmembership{Student Member,~IEEE,}
		Udaya Sampath K.P. Miriya Thanthrige\IEEEauthorrefmark{1}
,~\IEEEmembership{Student Member,~IEEE,}  Aly El Gamal,~\IEEEmembership{Senior~Member,~IEEE,} and Aydin Sezgin,~\IEEEmembership{Senior~Member,~IEEE}
\thanks{This   work   is   funded   by   the   Deutsche   Forschungsge-meinschaft  (DFG,  German  Research  Foundation)  Project-ID287022738 TRR 196 (S02 and S03)}
		\thanks{
		A.~M.~Ahmed, U.~S.~K.~P.~M.~Thanthrige, and A. Sezgin are with the Institute of Digital
		Communication Systems, Ruhr University Bochum, 44801 Bochum, Germany
		(e-mail: aya.mostafaibrahimahmad@rub.de; udaya.miriyathanthrige@rub.de;
		aydin.sezgin@rub.de).} 
		\thanks{ A. El Gamal is with the
Department of Electrical and Computer Engineering, Purdue University,
West Lafayette, IN, USA. e-mail: {
elgamala}@purdue.edu}
		\thanks{\textit{\IEEEauthorrefmark{1} These authors contributed equally to the work}}
}

\IEEEpubid{\begin{minipage}{\textwidth}\ \centering \textbf{1089-7798 \copyright\  2020 EU. Personal use is permitted. For any other purposes, permission must be obtained from the IEEE by emailing pubs-permissions@ieee.org. For more information https://www.ieee.org/publications/rights/index.html.}\end{minipage}
}

	\maketitle
	
	\begin{abstract}
		We present a MUSIC-based Direction of Arrival (DOA) estimation strategy using small antenna arrays, via employing deep learning for reconstructing the signals of a \emph{virtual} large antenna array. Not only does the proposed strategy deliver significantly better performance than simply plugging the incoming signals into MUSIC, but surprisingly, the performance is also better than directly using an actual large antenna array with MUSIC for high angle ranges and low test SNR values. We further analyze the best choice for the training SNR as a function of the test SNR, and observe dramatic changes in the behavior of this function for different angle ranges.
	\end{abstract}
	

	%
	\IEEEpeerreviewmaketitle
	
	\section{Introduction}
	Direction of arrival (DOA) estimation refers to estimating the direction of several target electromagnetic waves through receive antennas that form a \emph{sensor array}. DOA has a wide range of applications, e.g., radar, sonar, and wireless communications \cite{Vantrees}. Accurate DOA estimation can be achieved using large antenna arrays at the cost of increased hardware and computational complexity. However, multiple input multiple output (MIMO) radars with co-located antennas can offer virtual enlargement of the aperture at the receiver, using relatively few physical antennas.  This in turn significantly increases the maximum number of targets that could be detected, and enhances the angular resolution at a compact size, due to the fact that MIMO radars can transmit multiple probing signals, which can be correlated or uncorrelated  \cite{colocRadar,bliss}.  
	As an alternative approach, sparse array radars (also known as thin array radars) have been extensively studied in the literature and found to offer similar advantages as MIMO radars \cite{MIMOvsSparse}. The idea is to decompose a filled array into two sub-arrays, breaking the uniform spacing rule, hence achieving a larger aperture. By this means, it can offer similar target detection and angular resolution capabilities as the MIMO radar \cite{MIMO_SPARSE} with lower hardware cost. For this purpose, several array configurations were proposed in the literature \cite{MRA,nestedarrays,Coprime_Moeness}. 
	However, sparse arrays suffer from the effect of grating lobes due to the non-uniform spacing between the antennas, which leads to large estimation errors \cite{Bounds}. Furthermore, existing vector space DOA methods such as the MUltiple-SIgnal Classification (MUSIC) algorithm can not be directly applied, due to the rank deficiency of the correlation matrix. Hence, a spatial smoothing variant of MUSIC is proposed in  \cite{SSMUSIC_Coprime} for rank enhancement at the cost of increased computational cost. This is due to the fact that spatial smoothing must be performed for every DOA estimation.\\ 
	\IEEEpubidadjcol  \IEEEpubidadjcol 
	A comparison between MIMO and sparse array radars has been conducted in \cite{MIMOvsSparse}, where MIMO radars were found to be preferable when compactness is essential, since sparse arrays are characterized by their large aperture size. However, sparse arrays might be preferable when the hardware cost is the driving requirement, yet sparse arrays are not robust to sensor failures unlike uniform linear arrays (ULA) \cite{ComparisonSparse}, which could present an added challenge.
	In this work, we investigate a novel approach which enhances the angular resolution and target detection capacity while satisfying both low cost and compactness properties. We exploit the potential of deep neural networks to learn the mapping between two antenna arrays of different sizes. 	Specifically, we try to emulate the received signal of a large ULA using only a significantly smaller sub-array, without the need to increase the array aperture size, through training a deep neural network. This is followed by using the trained model for each received pulse to estimate the DOA via employing MUSIC without any further processing, thereby delivering the advantages of sparse arrays (i.e., hardware cost reduction) without increasing the aperture size, and without compromising accuracy. Also, no additional computational cost due to spatial smoothing is required. The contributions of this work can be summarized as follows: 
	\vspace{-1 mm}
	\begin{itemize}
    \item  DOA resolution enhancement of low antenna arrays using a deep neural network (DNN) that learns the mapping between received signals of two differently-sized antenna arrays. Surprisingly, the performance obtained by using actual high antenna arrays is not only tightly approximated, but exceeded at low SNR for high angle ranges. 
    \item Analysis of the best training SNR as a function of the test SNR, as well as in presence of test SNR uncertainty. Interestingly, the behavior of this function is observed to vary dramatically for different angle ranges.
    \item Analysis of the denoising capabilities of the proposed DNN. We attribute - based on experimental evidence - our approach's superior performance to the high antenna setup at high angle ranges and low SNR, to a denoising DNN functionality that effectively increases input SNR.
\end{itemize}

\section{System Model} \label{smodel}
Consider a co-located MIMO radar system with $M$ transmit (TX) antennas and $N$ receive (RX) antennas. Here, each transmit antenna with index $m$ transmits a narrow-band signal $s_m(t)$ with nondispersive propagation, that is perfectly orthogonal to the rest and consists of a train of $P$ non-overlapping pulses; each with duration $T$. 
For simplicity, we consider TX and RX antennas in a ULA configuration with antenna spacing of $d=\lambda/2$, where $\lambda$ is the wavelength. We further assume that there are $K$ targets in the scene. The radar cross section (RCS) - based on pulse $p$ - and the direction of arrival (DOA) of the $k$-th target are given by $\alpha_{k,p} \in \mathbb{C}$ and $\theta_k$, respectively. In this paper, the target RCS is modeled based on the Swerling model \RNum{2}, where it is fixed during the pulse interval $T$ and changes independently from one pulse interval to another \cite{swerling}. We define the transmit and receive steering vectors of the $k$-th target as $\mathbf{a}_t(\theta_k)=\left[1,e^{j{\rho}d \sin\theta_k},\hdots,e^{j{\rho}d (M-1)\sin\theta_k}\right]^T$ and $\mathbf{a}_r(\theta_k)=\left[1,e^{j{\rho}d \sin\theta_k},\hdots,e^{j{\rho}d (N-1)\sin\theta_k}\right]^T$, respectively. Here, $(\cdot)^T$ is the transpose, and  $\rho=\frac{2\pi}{\lambda}$. Here, we consider all targets as point targets. In that case, the received echo (reflected signal) from the target does not expand beyond the radar resolution cell \cite{barton1998radar}. The received signal $\mathbf{r}(t)\in \mathbb{C}^{N}$ after transmitting $P$ pulses is hence \cite{colocRadar},
 \begin{equation}
 \mathbf{r}(t)=\sum_{k=1}^{K}\sum_{p=1}^{P}\alpha_{k,p} ~\mathbf{a}_r(\theta_k)\mathbf{a}_t^T(\theta_k) 
 \mathbf{s}(t-pT) + \mathbf{n}(t),
 \end{equation}  
 where $\mathbf{n}(t) \in \mathbb{C}^{N}$ is independent and identically distributed (i.i.d) Gaussian noise with variance $\sigma^2$ and $\mathbf{s}(t)=[s_1(t),\hdots,s_m(t)]^T$. Next, at each receive antenna, the received signal $\mathbf{r}(t)$ is cross-correlated with $M$ matched filters corresponding to each transmit signal as given below
 \begin{gather}
 \mathbf{Z}_p=\int_{0}^{T} \mathbf{r}(t)~\mathbf{s}^{H}(t-pT) dt.
 \label{eq:received_pulse1}
  \end{gather}
Here, $(\cdot)^H$ is the conjugate transpose. Due to the perfect orthogonality of the transmit waveforms, $\mathbf{Z}_p$ in \eqref{eq:received_pulse1} is
  \begin{gather}
 \mathbf{Z}_p=\sum_{k=1}^{K}\sum_{p=1}^{P}\alpha_{k,p} ~\mathbf{a}_r(\theta_k)\mathbf{a}_t^T(\theta_k)\mathbf{I}
 +\int_{0}^{T}\mathbf{n}(t)\mathbf{s}^{H}(t-pT) dt.
 \label{eq:received_pulse}
 \end{gather}
 Here, $\mathbf{I}$ is an identity matrix. Next, we rearrange \eqref{eq:received_pulse} as
 \begin{equation}
 \mathbf{Y}=\mathbf{A}(\theta)\mathbf{X}+\mathbf{N},
 \label{eq:MIMO_receved}
 \end{equation}
 where $\mathbf{Y}~\in~\mathbb{C}^{MN\times P}$ is the receive signal and it is given as $\mathbf{Y}=[\vect(\mathbf{Z}_1),\hdots,\vect(\mathbf{Z}_P)]$. Here, $\vect(\mathbf{Z}_p)$ denotes the conversion of the matrix $\mathbf{Z}_p$ of \eqref{eq:received_pulse} into a column vector. The steering vector matrix $\mathbf{A}(\theta)$ is given by $[\mathbf{v}(\theta_1),\hdots,\mathbf{v}(\theta_K)]$, where $\mathbf{v}(\theta_k)= \mathbf{a}_t(\theta_k) \otimes \mathbf{a}_r(\theta_k)$.
 Further, the RCS matrix $\mathbf{X} \in \mathbb{C}^{K\times P}$ corresponding to all $K$ targets is given as $\mathbf{X}=[\mathbf{x}_1,\hdots,\mathbf{x}_P]$, with $\mathbf{x}_p=[\alpha_{1,p},\hdots,\alpha_{K,p}]^T$.
\begin{figure}
	\centering
	\includegraphics[width=1\linewidth]{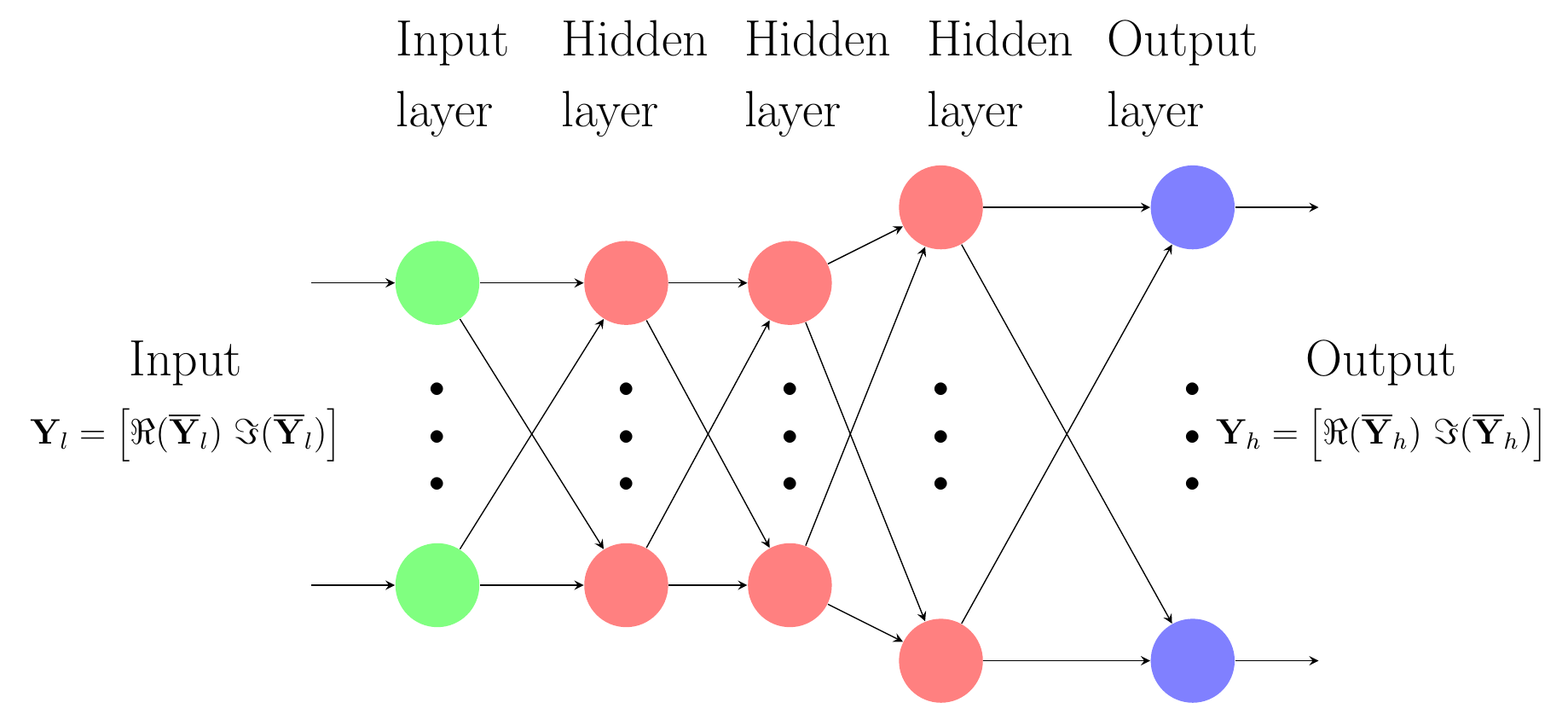}
	\caption{Training the Deep Neural Network.}
	\label{fig:nntraining}
\end{figure}
\begin{figure}
	\centering
	\includegraphics[width=1\linewidth]{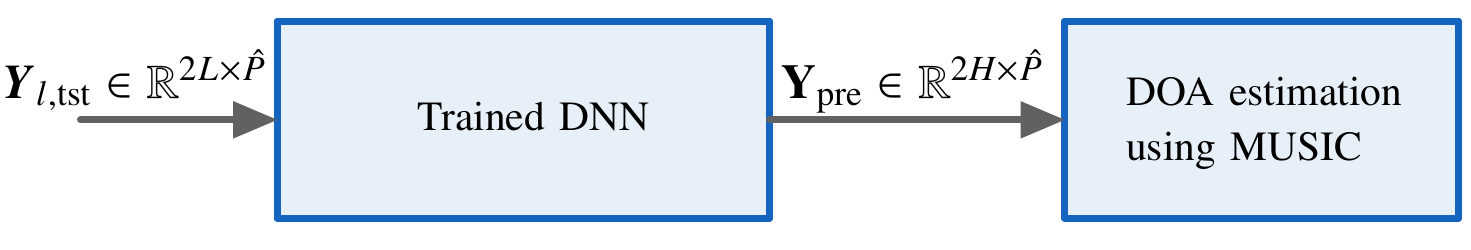}
	\caption{Antenna array reconstruction using the DNN.}
	\label{fig:nntesting}
\end{figure}
\section{Deep Learning Architecture}

Enlarging antenna array aperture enhances angular resolution capabilities, which in turn leads to better DOA estimation. 
Hence, we tackle the problem of mapping the received signal of two antenna setups of different sizes\footnote{The maximum number of targets ($K_{\max}$) that can be uniquely identified by a MIMO radar is given by $K_{\max}\in\left[{2(M+N)-5\over 3},{2MN\over 3}\right)$ \cite{4358016}. It is hence feasible to detect four targets for both considered low and high antenna setups. However, the accuracy of MUSIC suffers severe degradation as the SNR decreases, and this performance degradation can be significantly overcome by increasing the number of antennas, which we do through DNN emulation.}. A feedforward deep neural network (DNN) is proposed to learn the mapping between the received signals of low and high antenna setups.
Let $\overline{\mathbf{Y}}_l \in \mathbb{C}^{L\times P}$ and $\overline{\mathbf{Y}}_h \in \mathbb{C}^{H\times P}$ be the received signals of the low and high antenna setups as defined in (4), respectively. Here, $L=M_l N_l < H=M_h N_h$. $\overline{\mathbf{Y}}_h$ and $\overline{\mathbf{Y}}_l$ are then given as
  \begin{equation}
 \overline{\mathbf{Y}}_l=\mathbf{A}_{L}(\theta)\mathbf{X}+\mathbf{N},
 \label{eq:MIMO_recevedlow}
 \end{equation}
   \begin{equation}
 \overline{\mathbf{Y}}_h=\mathbf{A}_{H}(\theta)\mathbf{X}+\mathbf{N}.
 \label{eq:MIMO_recevedhigh}
 \end{equation}
Our approach is based on the hypothesis that in a complex environment, there is a non-linear relationship between both received signals corresponding to low and high antenna setups, which is a priori unknown due to the unknown locations of different targets. We hence train a DNN to learn this mapping in a data-driven fashion. The DNN consists of four fully connected layers, where the input layer is of dimension $L$, followed by three hidden layers of dimensions $L$, $L$, and $H$, respectively, and the output layer is of dimension $H$. The DNN architecture is shown in Fig. \ref{fig:nntraining}. Different DNNs with different configurations were tested to validate this selection. It was observed that the DNN with three hidden layers is the smallest DNN architecture that led  to a good performance, on average.  
Since the DNN is not designed for special processing of complex data, the input and output are defined as $\mathbf{Y}_l = \left[\Re (\overline{\mathbf{Y}}_l);~\Im (\overline{\mathbf{Y}}_l)\right]$, and $\mathbf{Y}_h = \left[\Re (\overline{\mathbf{Y}}_h);~\Im (\overline{\mathbf{Y}}_h)\right]$, where $\Re(\cdot)$, and $\Im(\cdot)$ denote the real and imaginary components, respectively. Both received  and reconstructed signals are normalized to lie between $0$ and $1$ through min-max normalization. ReLU is used as an the activation function for all the hidden layers. For the output layer, we tried both linear activation and ReLU, and then we chose the best performance for each experiment. The available dataset is divided into training, validation and testing, with split ratios of $60\%$, $20\%$ and $20\%$, respectively. Training takes place over a maximum of $150$ epochs with a batch size of $120$. For the training process, we used an Adam optimizer with the mean squared error loss function. In the testing phase, the DNN is tested using ${\mathbf{Y}}_{l,\mathrm{tst}} \in \mathbb{R} ^{2L\times \hat{P}}$, where it predicts ${\mathbf{Y}}_{\mathrm{pre}} \in \mathbb{R} ^{2H\times \hat{P}}$, as shown in Fig. \ref{fig:nntesting}. Here, $\hat{P}$ is the number of testing samples. DOA estimation is calculated from the predicted received signal $\overline{\mathbf{Y}}_{\mathrm{pre}} \in \mathbb{C} ^{H\times \hat{P}}$ through the MUSIC algorithm. The covariance matrix is calculated using $N_s$ snapshots as 
\begin{equation}
\label{cov}
\begin{aligned}
\mathbf{R}_{\mathrm{pre}}&=\E{[\overline{\mathbf{Y}}_{\mathrm{pre}}\overline{\mathbf{Y}}_{\mathrm{pre}}^H}]  = \mathbf{A}(\theta)\E{[\mathbf{X}\mathbf{X}^H}]\mathbf{A}^H(\theta)+\sigma^2 \mathbf{I},\\
&=\mathbf{U}_x\mathbf{\Lambda}_x\mathbf{U}_x^H+\mathbf{U}_n\mathbf{\Lambda}_n\mathbf{U}_n^H,
\end{aligned}
\end{equation}
 where $\E{[\cdot]}$ denotes the expected value, $\mathbf{U}_x$ and $\mathbf{U}_n$ are matrices containing the eigenvectors, which represent the signal and noise subspaces, respectively. $\mathbf{\Lambda}_x = \mathrm{diag}(\lambda_1,\hdots,\lambda_K)$ and $\mathbf{\Lambda}_n=\mathrm{diag}(\lambda_{K+1},\hdots,\lambda_{MN})$ contain the corresponding eigenvalues of the target and the noise, respectively. Hence, the expression of the MUSIC spectrum which provides the received signal energy distribution for all receive directions is given by $P_{MU}(\theta)=\left({\mathbf{v}^H(\theta)\mathbf{U}_n\mathbf{U}^H_n \mathbf{v}(\theta})\right)^{-1}$.
 For a comprehensive evaluation of our model performance, we define two metrics. First, we define the covariance matrix error as 
 \begin{equation}
    R_{e}=\left\lVert \mathbf{R}_{h,\mathrm{tst}}- \mathbf{R}_{\mathrm{pre}} \right\rVert_F\Big/ \left\lVert \mathbf{R}_{h,\mathrm{tst}} \right\rVert_F,
    \label{Re}
     \end{equation} 
 where $\left\lVert\cdot\right\rVert_F $ is the Frobenius norm, $\mathbf{R}_{h,\mathrm{tst}}=\E{[\overline{\mathbf{Y}}_{h,\mathrm{tst}}\overline{\mathbf{Y}}_{h,\mathrm{tst}}^H}] $, and $\overline{\mathbf{Y}}_{h,\mathrm{tst}}$ is the received signal of the high antenna setup during inference. The analysis of the covariance matrix of the received signal has a significant importance here as it is directly used to calculate the MUSIC spectrum. Second,
to evaluate the DOA estimation performance, the average mean squared error (MSE) over $Q$ trials is used as the performance metric. Here, $Q=\hat{P}/N_s$ . The MSE in radians is given by $\text{MSE}=\frac{1}{KQ}\sum_{q=1}^{Q} \sum_{k=1}^{K} (\hat{\theta}_{q,k}-\theta_{q,k})^2$. Here, the estimated and actual angles of the $k-$th target in the $q-$th trial are given as $\hat{\theta}_{q,k}$ and $\theta_{q,k}$, respectively. \\ The computational complexity of DNN training is governed by that of backpropagation, which is given by $\mathcal{O}(Pn(H^2+L^2))$. Here, $n$ is the number of training epochs and $\mathcal{O}(\cdot)$ represents the Big O notation for asymptotic computational complexity analysis. The test-time computational cost for a single trial which consists of $N_s$ snapshots is $\mathcal{O}(N_sH^2+N_sL^2)$, which is governed by inference cost along with MUSIC complexity, which is given by $\mathcal{O}(tH^2 +N_s H^2)$ where $t$ is the spatial grid search over the angles. Thus, total test-time computational cost is $\mathcal{O}(N_s(H^2+L^2)+tH^2)$.
\begin{figure*}[!h]
	\centering
	\includegraphics[width=1\linewidth]{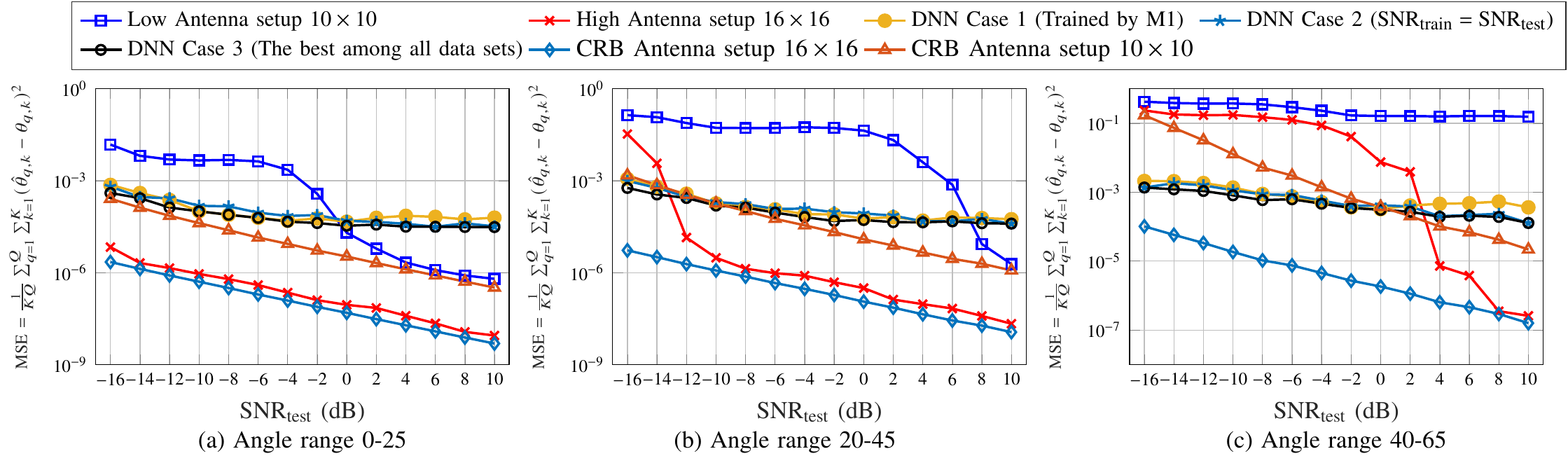}
	\caption{DOA estimation results and Cram\'{e}r–Rao Bounds.}
\label{Fig:CRB}
\end{figure*}
\begin{figure}[h]
	\centering
\includegraphics[width=1\linewidth]{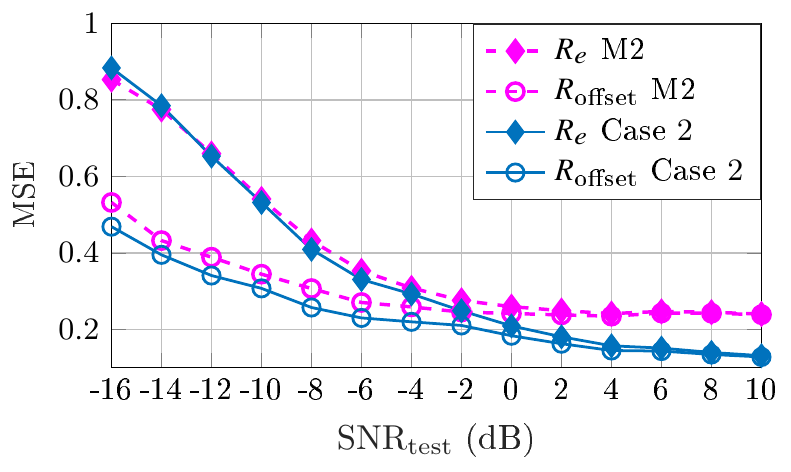}
\caption{MSE of covariance matrix of predicted received signal  compared to the actual high antenna setup at the same SNR and with SNR offset with ReLU output activation function.}
\label{RLangles}
\end{figure}
\begin{figure*}[h]
	\centering
\includegraphics[width=1\linewidth]{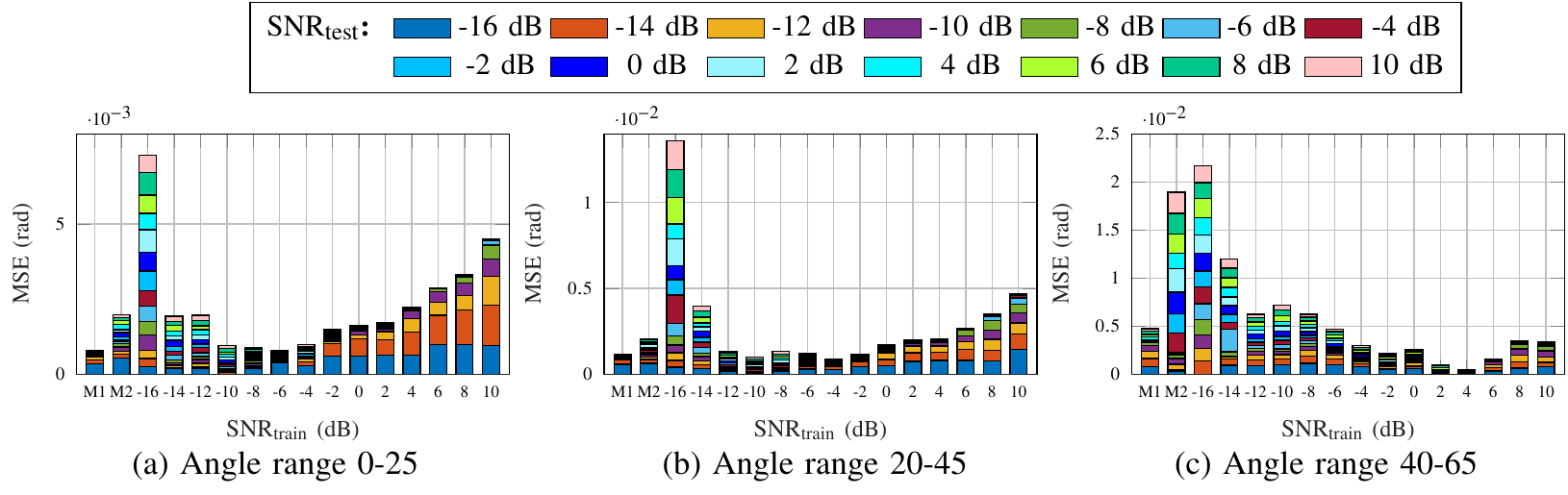}
	\caption{DOA estimation comparison of the DNN based signal prediction by training at a single SNR.} \label{fig:all_bar}
	\end{figure*}
\begin{figure*}[h]
	\centering
\includegraphics[width=1\linewidth]{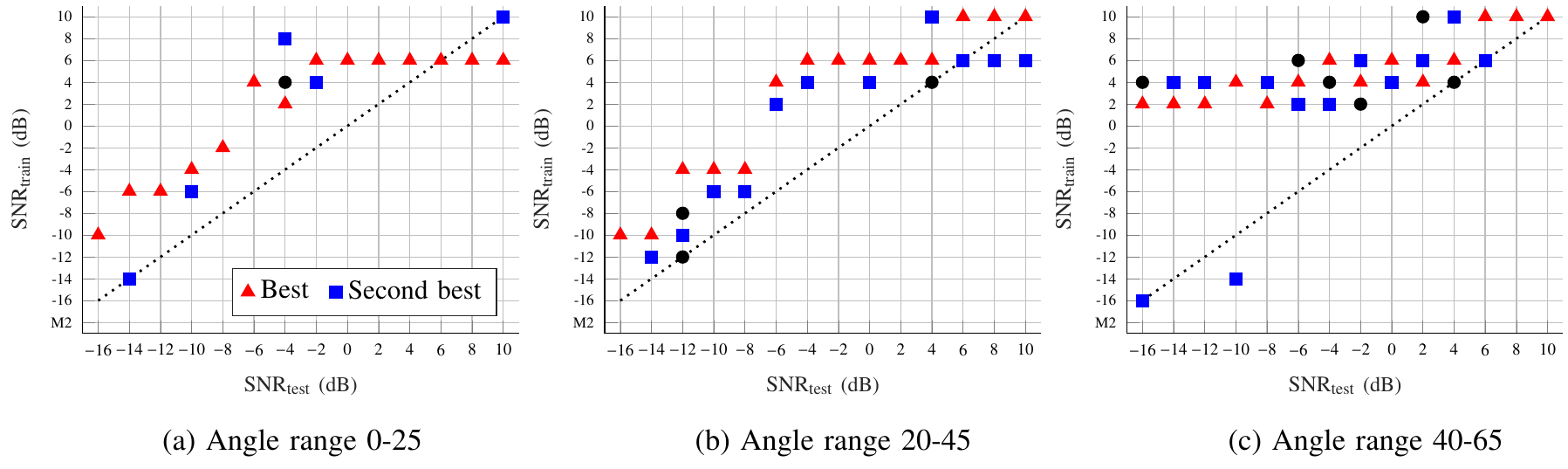}
	\caption{The best training SNR selection for different test SNRs.} \label{fig:grid_10}
\end{figure*}
%
	\begin{table}[!t]
		\centering
		\begin{tabular}{|l|l|l|l|}
			\hline
			$N_l$ $=$ $M_l$                                                                                 & $10$  & $N_h$ $=$  $M_h$                                                                                & $16$    \\ \hline				
			Angle grid 1  & $0:25$ &  Angle grid 2  & $20:45$ \\ \hline
			Angle grid 3  & $40:65$ &  Number of targets ($K$) & $4$ \\ \hline
		\end{tabular}
		\caption{Simulation Parameters}
		\label{table_1}
		\vspace{-0.5cm}
	\end{table}

\section{Simulation results}
To train the DNN, we use a GPU server with $32$ GB memory and a single NVIDIA Quadro RTX $5000$. 
The simulation parameters are listed in Table \ref{table_1}.
We consider three angle ranges, which span the scope of the incident signal, however similar results were obtained for others. Each range is chosen to span $25$ degrees to place the targets. Also, here we set the minimum spatial distance of five degrees between two targets to ensure the best spatial resolution of the actual large antenna setup of $16$ $\times$ $16$ antennas. Different training sets are considered in the training phase. Specifically, we consider $16$ datasets\footnote{Source code is available for download at https://gitlab.com/miriyugl/doa-with-dnn-via-emulation-of-antenna-arrays.} with different SNR combinations. Here, two datasets (M1 and M2) contain data with a mix of SNR values while the other $14$ datasets involve only a single SNR each. M1 and M2 contain equal percentage of data samples from each training SNR ($\mathrm{SNR}_\mathrm{train}$) ranging from $-16$ dB to $10$ dB with a step size of $2$ dB. The only difference in constructing these two datasets is the size, as M1 consists of $840000$ and M2 consists of $60000$ samples for both training and validation. Furthermore, each of the other $14$ datasets consists of $60000$ samples for training and validation. For the testing phase, we estimate the DOA for $15000$ samples in the testing SNR ($\mathrm{SNR}_\mathrm{test}$) range of $-16 \dots 10$ dB with the same training step size. Here, we set the number of snapshots $N_s$ as $150$, resulting in a number of trials $Q=100$.\\
 The average MSEs for different training datasets are shown in Fig. \ref{Fig:CRB}. In this figure, DOA estimation using the predicted signal of the DNN is compared with the DOA estimation obtained by directly using the signals obtained from the actual low and high antenna setups. Three cases for the DNN prediction task are explored in this figure:
\vspace{-0.1cm}
\begin{itemize}
 \item Case 1: Training the DNN with the M1 dataset (i.e., mix of SNR values).
\item Case 2: Training the DNN with the same SNR as that used in testing (e.g., using the DNN model trained with $\mathrm{SNR}_\mathrm{train}=$ $-16$ dB at $\mathrm{SNR}_\mathrm{test}=$ $-16$ dB).
\item Case 3: Selecting the lowest MSE of DOA estimation achieved across all $16$ data sets for each testing SNR (e.g., using the DNN model trained with $\mathrm{SNR}_\mathrm{train}=-10$ dB leads to the lowest MSE at $\mathrm{SNR}_\mathrm{test}=-16$ dB for the angle range of $20-45$ degrees).
\end{itemize}

Fig. \ref{Fig:CRB} shows that the  predicted  signal typically leads to better performance than directly using the signal of the low antenna setup (10$\times$10), specially in the low SNR regime and for high angle ranges. However, there is one exception to this statement. As shown in  Fig. \ref{Fig:CRB}(a) and (b), where the  low  antenna  setup  performs slightly  better  compared to  the  DNN  prediction in high SNR regimes.  
 This is due to the fact, that both high and low received signals have similar DOA performance in high SNR, where the MSE decreases dramatically in both cases. We believe that the inferior DNN performance in this case can be attributed to overfitting as the training loss is lower than the validation loss.  
 Fig. \ref{Fig:CRB} also demonstrates that training and testing with the same SNR closely follows the best achievable performance, and hence highlighting the impact of knowing the test SNR value and choosing the simple strategy of training at only that value. 
In addition, Fig. \ref{Fig:CRB} demonstrates the difference in behavior among different angle ranges. More specifically, the performance of the DOA estimation obtained from using the signals corresponding to the actual antenna setups becomes worse for higher angle ranges. We believe that this is due to the loss of spatial resolution of the ULA as the target directions shift to the endfire direction of the antenna array (i.e. $|\theta_k|\geq 60$). This is due to the fact that in this range, the beam sharpness reduces remarkably as the effective array aperture decreases towards those directions \cite{Balanis}.\\
Interestingly, DOA estimation using our DNN-emulated signal outperforms the one generated using the actual high antenna setup at low SNR and high angle ranges. A possible explanation of this behavior is that, while pursuing improvement in generalization performance, the DNN performs denoising to the received signal. We further examine this hypothesis by evaluating $R_e$ as defined in \eqref{Re}. Then, we compare the predicted signal with the actual received signal at a certain SNR offset. Hence, we define $R_{\mathrm{offset}}$ as 
\begin{equation}
    R_{\mathrm{offset}}=\left\lVert \mathbf{R}_{ho,\mathrm{tst}}- \mathbf{R}_{\mathrm{pre}} \right\rVert_F \Big/ \left\lVert \mathbf{R}_{ho,\mathrm{tst}} \right\rVert_F,
    \label{eq:Roffset}
     \end{equation}
     where $\mathbf{R}_{ho,\mathrm{tst}}=\E{[\overline{\mathbf{Y}}_{ho,\mathrm{tst}}\overline{\mathbf{Y}}_{ho,\mathrm{tst}}^H}]$, and $\overline{\mathbf{Y}}_{ho,\mathrm{tst}}$ is the actual received signal at a certain SNR offset (e.g., if $\overline{\mathbf{Y}}_{h,\mathrm{tst}}$ and $\overline{\mathbf{Y}}_{\mathrm{pre}} $  are evaluated using $\mathrm{SNR}_{\mathrm{test}}=$ $-16$ dB, then $\overline{\mathbf{Y}}_{ho,\mathrm{tst}}$ is evaluated using  $\mathrm{SNR}_{\mathrm{test}}=$ $-8$ dB with an offset of $8$ dB).
     In Fig. \ref{RLangles}, we plot $R_e$ and $ R_{\mathrm{offset}}$ using the training datasets of M2 and case 2 with offset values of $8$ and $12$ dB, respectively. Those offset values are chosen based on the observed performance corresponding to both cases. 
     Fig. \ref{RLangles} shows that $ R_{\mathrm{offset}}$ has much lower values compared to $R_e$ in both cases. 
     That signifies the statistical similarity between the predicted signal of the DNN and the less noisy version of the actual received signal of the high antenna setup. Further, as the SNR increases, $ R_{\mathrm{offset}}$  and $R_e$ converge to the same value. This underlines the validity of the hypothesis that the DNN denoises the received signal.
     \vspace{-0.2cm}
\subsection{Cram\'{e}r–Rao bound (CRB) Analysis.}\label{crb}
To further assess the performance of our approach, the Cram\'{e}r–Rao bound (CRB) of unbiased DOA estimation of MIMO radar is calculated and derived as in \cite{CRBMIMO,CRBrev1}. 
\begin{equation*}
CRB(\theta)={\sigma^{2}\over 2N_s}\left\{{\rm Re}\left[{\bf X}^H{\bf A}_{e}^{H}\left(I-{\bf A}\left({\bf A}^{H}{\bf A}\right)^{-1}{\bf A}^{H}\right){\bf A}_{e}{\bf X}\right]\right\}^{-1}.
\label{eq:crb}
\end{equation*}
Here, $\bf {A}_{e}$ represents first derivative information of the steering vector matrix $\mathbf{A}=\mathbf{A}(\theta)$ as given by 
${\bf {A}_{e}=\bf {A}_{e}(\theta)=\left[{\partial {\bf v}(\theta_{1})\over \partial\theta_{1}},{\partial {\bf v}(\theta_{2})\over \partial\theta_{2}},\cdots,{\partial {\bf v}(\theta_{K})\over \partial\theta_{K}}\right] }$.  Fig. \ref{Fig:CRB} shows the CRB of the DOA estimation of low and high antenna setups. It can be seen that the DOA estimation of the DNN-predicted signal approaches the CRB of the low antenna setup, specifically in the low SNR regime. Further, for higher angle ranges at low SNR, the MSE of the DNN-based DOA estimation is lower than the CRB of the low antenna setup. We believe that this is due to the side information benefit from the high antenna setup training signals, as well as the denoising effect.
\subsection{What is the best training SNR values?}\label{snr_range}
We first investigate the performance when training with a single SNR value across all testing SNR values. Fig. \ref{fig:all_bar} shows the cumulative average MSEs of the DOA estimation over all testing SNRs for each training SNR. 
Note that the shortest bar corresponds to the training SNR which provides the lowest cumulative MSE over all testing SNR values. We observe that the training set M1 consistently provides a low cumulative MSE. However, it may be difficult in practice to acquire - and train with - a large dataset due to latency and computational constraints. Interestingly, the best training SNR value, in terms of cumulative MSE, shifts from low to high as we move towards higher angle ranges. Further, perhaps counterintuitively, training with high SNR values can lead to mild performance for low angle ranges in presence of uncertainty about the testing SNR.\\ Further analysis is conducted to elaborate the relationship between the training and test SNR, and the results are shown in Fig. \ref{fig:grid_10}. 
Consider the average MSE for each pair of training and testing SNR values. With respect to that metric, in 
Fig. \ref{fig:grid_10}, \addlegendimageintext{mark=triangle*, mark size=4pt, color=red, fill=red} and \addlegendimageintext{mark=square*, mark options={}, mark size=3pt, color=blue, fill=blue} represent the best and the second best training SNRs for a particular testing SNR, respectively. Further, \addlegendimageintext{mark=*, mark options={}, mark size=3pt, color=black, fill=black} represents other training SNRs which deliver an average MSE within 10\% of the best. 
We observe that positive training SNR values are in general more favorable, specially as we move towards higher angle ranges and higher SNR values. However, when comparing with the results in Fig. \ref{fig:all_bar}, we conclude that knowledge of the test SNR favors higher training SNR values, while a significant uncertainty about the test SNR favors lower training SNR values, specially for lower angle ranges. 
\section{Conclusion}
We introduced a novel strategy that employs deep learning for emulating large antenna arrays, and demonstrated how it boosts the accuracy of MUSIC for Direction Of Arrival (DOA) estimation. Multiple observations - of practical significance - were drawn from the obtained results. Most notably, we highlighted how the emulated array leads to superior performance than an actual antenna array with the same number of antennas for high angle ranges and low SNR values, probably due to the denoising abilities of deep neural networks. Further, the effectiveness of training at low SNR values in presence of uncertainty about the test SNR was demonstrated, especially for low angle ranges. Finally, we investigated the best training SNR values as a function of the test SNR, and particularly noted the shift in ideal training SNR values from low to high as we move towards higher angle ranges and higher test SNR values.  
\appendices

	
\bibliographystyle{IEEEtran}
\bibliography{refs}

\end{document}